# Hit Selection Using SSMD-Based Machine Learning Performance Metrics in High-Throughput Screening Assays

**Xiaohua Douglas Zhang**

Department of Biostatistics, College of Public Health, University of Kentucky, Lexington, KY40536, USA

Tel.: 8595623373; Email: douglas.zhang@uky.edu

**Abstract**

High-throughput screening (HTS) assays are central to early-stage drug discovery but are often limited by extreme data sparsity, as primary screens typically use only a single replicate per test substance. This sparsity makes conventional machine-learning performance metrics, such as sensitivity, specificity, and area under the receiver operating characteristic curve (AUROC), difficult to estimate empirically because they require adequately sized labeled samples. Here, we introduce a model-based framework that derives these classification metrics from the strictly standardized mean difference (SSMD), a well-established HTS effect-size parameter. Under a Gaussian equal-variance assumption, we derive closed-form relationships linking SSMD to Youden-optimal sensitivity and specificity, and sensitivity at a preset specificity, yielding explicit estimators and exact confidence intervals from the noncentral t-distribution, even under single-replicate designs. Unlike classical statistical power, which approaches 1 as sample size grows regardless of how small the true non-zero difference between group means is, the SSMD-derived sensitivity converges to a finite population value that reflects the true degree of separation between two groups, making it a more meaningful and stable performance measure for hit selection. We demonstrate the utility of this framework in a hepatitis C virus primary siRNA screen comprising approximately 22,000 single-replicate measurements, showing that SSMD, AUROC, and sensitivity-based thresholds yield equivalent and interpretable hit sets. This work bridges classical HTS statistics and machine-learning evaluation theory, providing a statistically principled, reproducible way to estimate classification performance in ultra-low-replication screening workflows.

## 1. Introduction

HTS is a core technology in drug discovery and functional genomics, enabling systematic evaluation of large libraries of perturbations—including small molecules, siRNAs, sgRNAs, peptides, antibodies, and proteins—against biological targets (Shalem et al., 2015). Advances in automation and assay miniaturization allow modern HTS platforms to evaluate hundreds of thousands to millions of test substances per campaign, substantially accelerating early-stage discovery (Arkin & Wells, 2004). Most HTS studies adopt a two-stage design: a primary screen, typically conducted with a single replicate per test substance to maximize throughput, followed by a confirmatory screen in which a small subset of candidates is reassessed using two to four replicates to improve reproducibility and reduce false positives (Zhang, 2011b).

Common heuristics for HTS hit selection—such as percent inhibition, fold change, and mean difference—are intuitive but ignore variability and are poorly calibrated for decision-making under uncertainty. More formal methods, including Z-scores, their robust variants such as Z* and B-scores, and p-values from t-tests, improve robustness but do not directly quantify classification performance and suffer from well-known limitations in high-throughput settings (Zhang, 2011a). SSMD was introduced to address these limitations and has therefore emerged as a preferred effect-size metric because it integrates signal magnitude and variability and applies to both single- and multiple-replicate designs (Zhang, 2007a, 2007b). While SSMD provides interpretable thresholds and is widely used in practice (Han et al., 2022; Han et al., 2025; Hao et al., 2022; Jiang et al., 2021; Lim, 2023; Pellattiero et al., 2025; Rossiter et al., 2021; Yin et al., 2025; Zhang, 2007a, 2007b; X. D. Zhang, 2008; Zhang, 2011b; Zhang, 2025;

Zhang et al., 2011; Zhang et al., 2020; Zhou et al., 2008; Zhou et al., 2014), its established relationship with AUROC has mainly supported quality control (Zhang, 2025) and has not directly yielded key machine-learning metrics such as sensitivity and specificity.

Machine learning offers a principled framework for formalizing HTS hit selection as binary classification under severe data scarcity. Performance metrics such as sensitivity, specificity, and AUROC provide explicit control over the trade-off between discovery of true actives and propagation of false positives into costly downstream validation (Stokes et al., 2020). Despite their relevance, these metrics are rarely used in HTS workflows because non-parametric estimation requires sufficiently large labeled datasets, a requirement fundamentally violated in single-replicate primary screens and barely satisfied in confirmatory studies (Zhang, 2025). This limitation reflects a broader challenge in machine learning: how to meaningfully evaluate and optimize classification performance when labeled data are extremely sparse.

In this work, we address this challenge by introducing a model-based framework that links SSMD to machine-learning performance metrics through an explicit analytical relationship with AUROC. Under the assumption of normally distributed assay measurements with equal variance—a standard assumption in HTS quality control and effect-size modeling (Zhang, 2011b)—we derive closed-form expressions for sensitivity and specificity, as functions of SSMD-based decision thresholds. This approach enables performance-aware hit selection even in single-replicate primary screens, where empirical estimation of classification metrics is infeasible. By establishing a formal connection among SSMD, AUROC, Sensitivity and Specificity, our method bridges classical HTS statistics with machine-learning evaluation theory under extreme data sparsity.

# 2. Methods

## 2.1. Problem Setup and Assumptions

HTS experiments are typically performed in standardized microtiter plates—commonly 96-, 384-, or 1536-well formats—where each well serves as an independent reaction or assay condition (Zhang, 2011b). With the assumption that the variance of a test substance equals that of the negative reference, we derive the relationship among SSMD, AUROC, specificity, and sensitivity under the condition that both the test substance and the negative reference are normally distributed with equal variance within each plate in an HTS study.

We formalize HTS hit selection as a binary classification problem under extreme data sparsity. For each test substance, a single noisy observation is available in the primary screen, while negative reference measurements are replicated multiple times per plate.

Let $Y_0 \sim N(\mu_0, \sigma_0^2)$ denote the response of the negative reference (class 0) and $Y_1 \sim N(\mu_1, \sigma_1^2)$ denote the response of a test substance (class 1), where $Y_1 \perp Y_0$.

Hit selection is directional. We define

- up-regulated hits if $\mu_1 \geq \mu_0$
- down-regulated hits if $\mu_1 < \mu_0$.

SSMD is defined as (Zhang, 2007b), $\beta = \frac{\mu_1 - \mu_0}{\sqrt{\sigma_0^2 + \sigma_1^2}}$

In the condition of equal variance (i.e., $\sigma_0^2 = \sigma_1^2 = \sigma^2$),

$$\beta = \frac{\mu_1 - \mu_0}{\sqrt{2}\sigma} \tag{1}$$

Hereafter, we model HTS hit selection as a binary classification problem under extreme data sparsity, assuming normally distributed test and control responses with equal variance within each plate. Under this framework, we derive closed-form relationships linking SSMD to AUROC, Youden-optimal sensitivity and specificity, and sensitivity at a preset specificity under the setting for HTS hit selection, as shown in the Appendix. The derived results are shown in Table 1.

These results eliminate the need for empirical ROC construction. They also enable direct estimation of classification performance from SSMD estimates and exact confidence intervals are obtained via the noncentral *t*-distribution and mapped analytically to AUROC, sensitivity, and specificity as shown in the section below.

## 2.4. Estimation and Inference

In practice, $\mu_0, \mu_1, \sigma^2$ are unknown. Under normality with equal variance, the two sample *t*-statistic

$$T = \frac{\overline{Y_1} - \overline{Y_0}}{\sqrt{\frac{2}{\nu H}\left((n_0-1)s_1^2+(n_1-1)s_1^2\right)}} \sim \text{noncentral } t\left(\nu, \sqrt{H}\beta\right), \text{ where } \nu = n_0 + n_1 - 2 \text{ and } H = \frac{2}{\frac{1}{n_0}+\frac{1}{n_1}},$$

$\overline{Y_0}$ and $s_0^2$ are the sample mean and variance of $n_0$ measured values of the negative reference in a plate, respectively, and $\overline{Y_1}$ and $s_1^2$ are the sample mean and variance of $n_0$ measured values of a test substance in a plate, respectively. Set $s_1^2 = 0$ if $n_1 = 1$.

When $n_0 + n_1 \geq 4$, the uniformly minimum variance unbiased estimator (UMVUE) of SSMD (Zhang, 2008) is

$$\hat{\beta} = \frac{\bar{Y}_1 - \bar{Y}_0}{\sqrt{\frac{2}{K}\left((n_0-1)s_0^2+(n_1-1)s_1^2\right)}} = \sqrt{\frac{K}{\nu H}}T;\ K = 2\left(\frac{\Gamma\left(\frac{\nu}{2}\right)}{\Gamma\left(\frac{\nu-1}{2}\right)}\right)^2 \quad (6).$$

The noncentral t-distribution $T$ in Formula (5) enables exact confidence intervals $(\beta_{\alpha/2}, \beta_{1-\alpha/2})$ for SSMD where $\beta_{\alpha}$ is the value such that $\Pr(\text{noncentral } t(\nu, \sqrt{H}\beta_{\alpha}) \leq T_{obs}) = \alpha$.

Exact confidence intervals follow from the noncentral t-distribution and are mapped to AUROC, sensitivity and/or specificity based on the results shown in Table 1. Estimated values and confidence intervals are summarized in Table 2.

*Table 2*. SSMD-Based Estimation of AUROC, Sensitivity, and Specificity for HTS hit selection.

| Metric | Quantity | Up-regulation | Down-regulation |
|---|---|---|---|
| **SSMD** | estimate | $\hat{\beta}$ | $\hat{\beta}$ |
| | $1-\alpha$ CI | $\left(\beta_{\frac{\alpha}{2}}, \beta_{1-\frac{\alpha}{2}}\right)$ | $\left(\beta_{\frac{\alpha}{2}}, \beta_{1-\frac{\alpha}{2}}\right)$ |
| **AUROC** | estimate | $\boldsymbol{\Phi}(\hat{\beta})$ | $\boldsymbol{\Phi}(-\hat{\beta})$ |
| | $1-\alpha$ CI | $\left(\boldsymbol{\Phi}\left(\beta_{\frac{\alpha}{2}}\right), \boldsymbol{\Phi}\left(\beta_{1-\frac{\alpha}{2}}\right)\right)$ | Symmetric |
| **Youden-Optimal** Sens/Spec | Threshold | $\frac{\bar{Y}_0 + \bar{Y}_1}{\mathbf{2}}$ | $\frac{\bar{Y}_0 + \bar{Y}_1}{\mathbf{2}}$ |
| | Estimate | $\boldsymbol{\Phi}\left(\frac{\hat{\beta}}{\sqrt{\mathbf{2}}}\right)$ | $\boldsymbol{\Phi}\left(-\frac{\hat{\beta}}{\sqrt{\mathbf{2}}}\right)$ |
| | $1-\alpha$ CI | $\left(\boldsymbol{\Phi}\left(\frac{\beta_{\frac{\alpha}{2}}}{\sqrt{\mathbf{2}}}\right), \boldsymbol{\Phi}\left(\frac{\beta_{1-\frac{\alpha}{2}}}{\sqrt{\mathbf{2}}}\right)\right)$ | Symmetric |
| **Preset Spec Sp** Sens | Threshold | $\bar{\bar{Y}}_0 + s_0 \cdot \Phi^{-1}(\mathrm{Sp})$ | $\bar{\bar{Y}}_0 - s_0 \cdot \Phi^{-1}(\mathrm{Sp})$ |
| | Estimate | $\boldsymbol{\Phi}\left(\sqrt{\mathbf{2}}\hat{\beta} - \boldsymbol{\Phi}^{-\mathbf{1}}(\mathbf{Sp})\right)$ | $\boldsymbol{\Phi}\left(-\sqrt{\mathbf{2}}\hat{\beta} - \boldsymbol{\Phi}^{-\mathbf{1}}(\mathbf{Sp})\right)$ |
| | $1-\alpha$ CI | mapped from $\left(\beta_{\frac{\alpha}{2}}, \beta_{1-\frac{\alpha}{2}}\right)$ | Symmetric |

Note:

- $\Phi(\cdot)$ is the cumulative distribution function of the standard normal distribution.
- $\hat{\beta} = \sqrt{\frac{K}{\nu H}} T_{obs}$ where $T_{obs}$ is the observed two-sample *t*-statistic.
- $H = \frac{2}{\frac{1}{n_0} + \frac{1}{n_1}}$, $\nu = n_0 + n_1 - 2$, $K = 2\left(\frac{\Gamma\left(\frac{\nu}{2}\right)}{\Gamma\left(\frac{\nu-1}{2}\right)}\right)^2$.

The exact confidence interval for $\beta$ is obtained from the noncentral *t*-distribution and mapped to AUROC and sensitivity using Theorems 1–3.

## 3. Application

We illustrate how the proposed framework integrates classical HTS statistics with machine-learning performance metrics to enable principled hit selection under ultra–low-replication settings.

The method is applied to a primary siRNA high-throughput screen comprising approximately 22,000 siRNAs distributed across 97 plates with 384 wells per plate, designed to identify host factors associated with hepatitis C virus (HCV) replication. Each siRNA was measured once (single replicate), consistent with standard primary HTS practice (Zhang et al., 2008). Within each plate, 16 wells contained positive controls, 16 wells contained negative controls, and the remaining 320 wells corresponded to test siRNAs.

Data preprocessing followed standard HTS normalization procedures. Raw responses were log2-transformed, after which percent inhibition was computed as , $\frac{y-\bar{X}_-}{\bar{X}_+-\bar{X}_-} \times 100$, where y is the log2-transformed value in a well, $\bar{X}_+$ is the sample mean of positive control and $\bar{X}_-$ is the 5% trimmed sample mean of all the tested siRNAs wells as most of the tested siRNAs should have no effect. Plate-level quality control was performed using the estimated SSMD between positive and negative controls, requiring it to exceed the critical value corresponding to a true SSMD of 3 (Zhang, 2025). Eighty-three plates passed this criterion and were retained for downstream analysis.

For each retained plate, we computed SSMD, mean difference in percent inhibition relative to the negative control, and the proposed model-based estimates of AUROC, sensitivity, and specificity. Sensitivity was evaluated both at the optimal Youden index and at a fixed specificity of 0.975, reflecting a conservative false-positive rate appropriate for primary screening. Results are visualized using plate-well series plots (Zhang, 2011b), as shown in Figure 1.

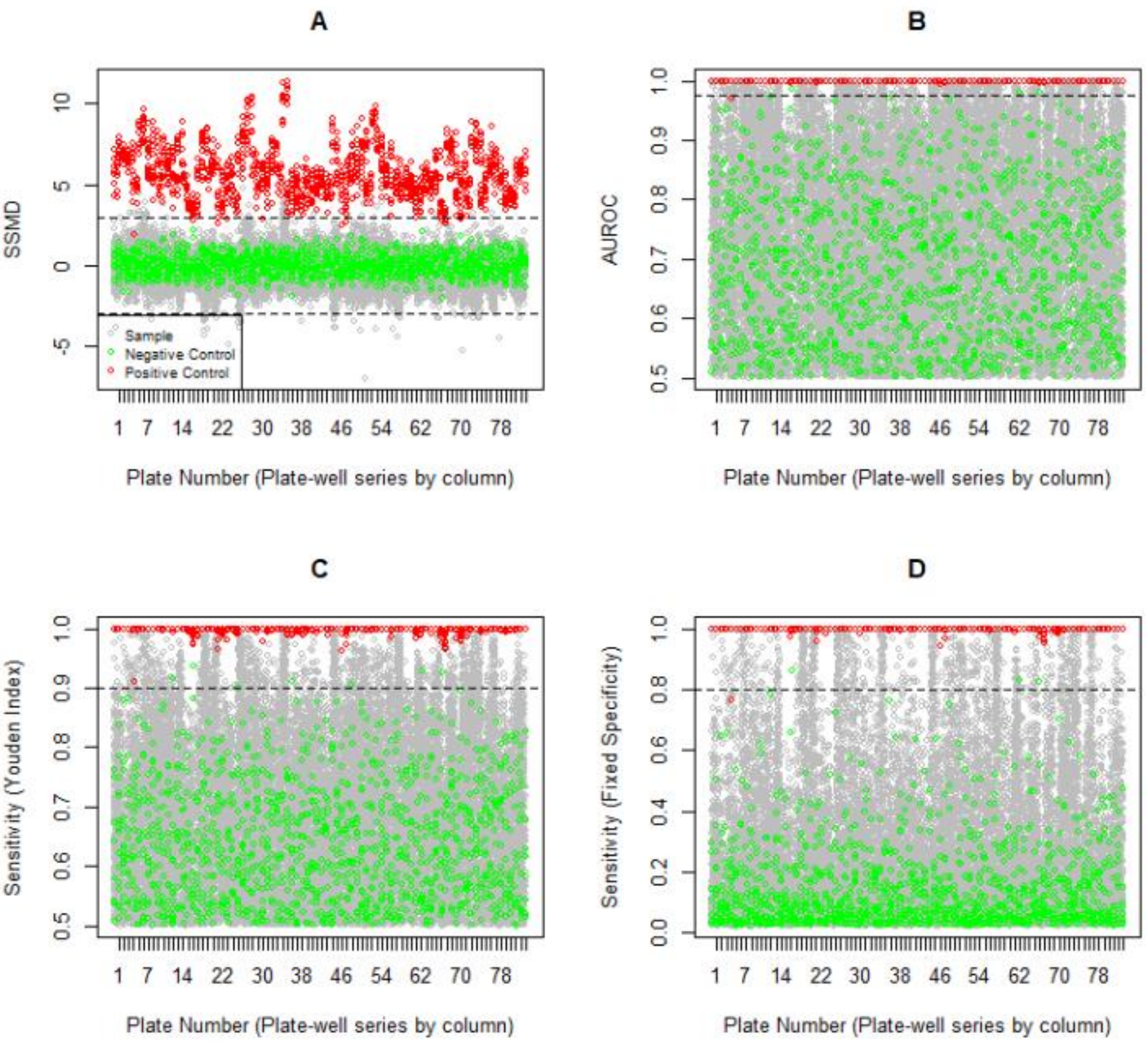

*Figure 1.* Plate–well series plots summarizing hit-selection metrics for the 83 plates that passed quality control in the HCV primary high-throughput screening study. Panel A shows the estimated SSMD; Panel B shows the corresponding AUROC derived from SSMD; Panel C shows sensitivity and specificity estimated by optimizing the Youden index; and Panel D shows sensitivity estimated at a fixed specificity of 0.975. Each point represents an individual well, enabling direct comparison of classical HTS metrics and their machine-learning performance interpretations across plates.

These results enable hit selection under four alternatives but mathematically linked criteria: SSMD, AUROC, sensitivity at the optimal Youden index, and sensitivity at a fixed specificity.

Using SSMD, siRNAs with estimated SSMD $\geq 1.96$ were classified as inhibition hits and those with SSMD $\leq -1.96$ as activation hits (Figure 1A), yielding 269 inhibition hits and 793 activation hits. Under the analytical relationships derived in this work, these thresholds correspond to AUROC values of at least $\Phi(1.96) = 0.975$, sensitivity and specificity at the optimal Youden index of at least $\boldsymbol{\Phi}\left(\frac{1.96}{\sqrt{2}}\right) = \mathbf{0.917}$, and sensitivity of at least $\boldsymbol{\Phi}\left(\sqrt{\mathbf{2}} \times 1.96 - \boldsymbol{\Phi}^{-\mathbf{1}}(\mathbf{0.975})\right) = \mathbf{0.792}$ when specificity is fixed at 0.975.

Applying AUROC directly with a cutoff of 0.975 (Figure 1B) yields an identical set of hits, demonstrating the equivalence between SSMD-based and AUROC-based selection under the assumed model. Alternatively, selecting hits using sensitivity and specificity jointly via the Youden index with a cutoff of 0.90 (Figure 1C) identifies 355 inhibition hits and 1,101 activation hits. Because a specificity of 0.90 allows a relatively high false-positive rate, a more conservative strategy fixes specificity at 0.975 and selects siRNAs with sensitivity exceeding 0.80 (Figure 1D), resulting in 260 inhibition hits and 748 activation hits.

To support practical decision-making, we further visualize effect size and classification performance jointly. The dual-flashlight plot (Figure 2A) displays mean percent inhibition against SSMD, while volcano-style plots combine percent inhibition with AUROC (Figure 2B), sensitivity at the Youden index (Figure 2C), and sensitivity at fixed specificity (Figure 2D). These visualizations demonstrate how traditional HTS effect-size measures can be interpreted through machine-learning performance metrics, enabling transparent and statistically grounded hit selection even in single-replicate primary screens.

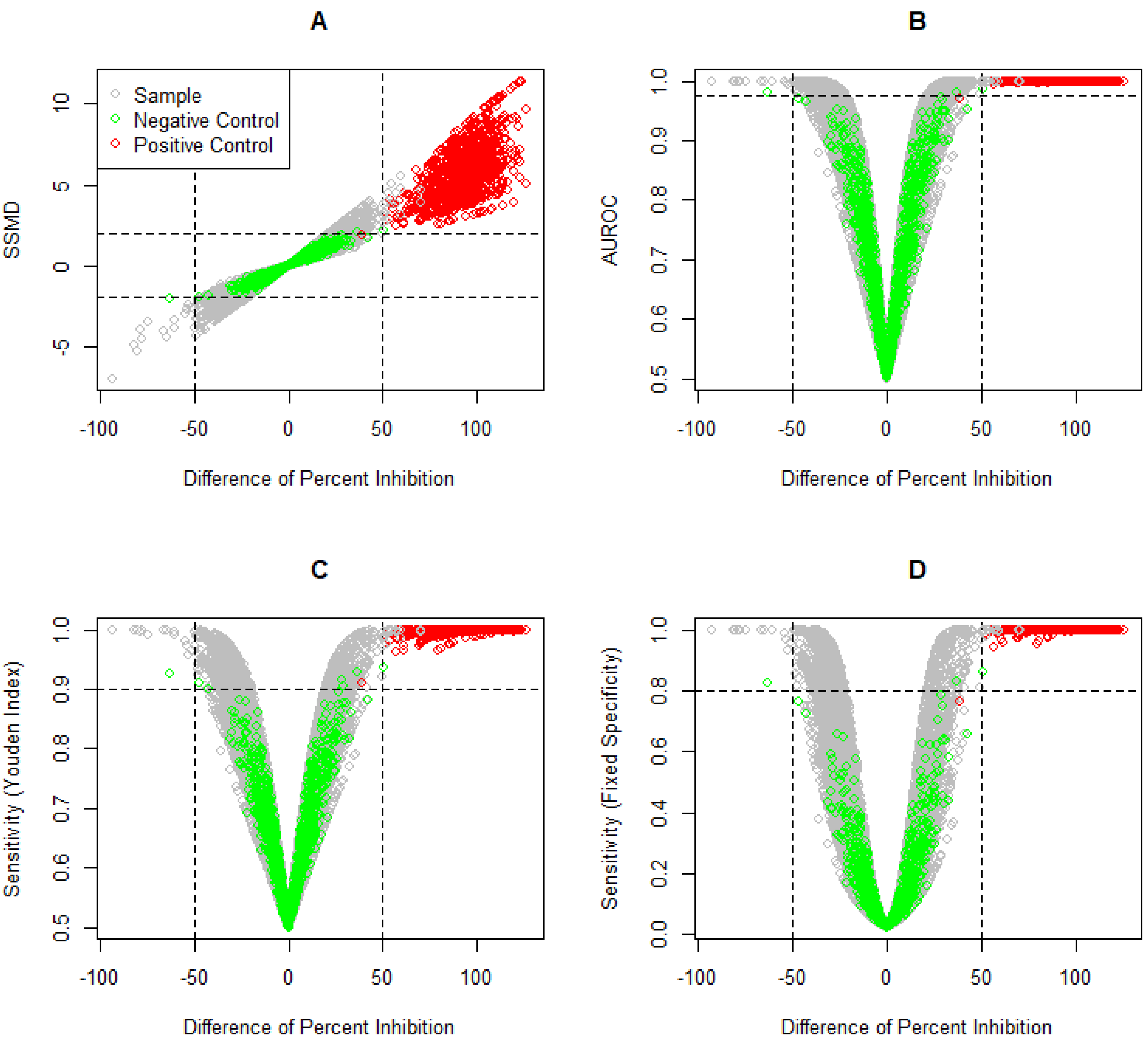

*Figure 2.* Scatter plots illustrating the joint use of effect size and variability-aware performance metrics for hit selection. The x-axis shows the difference in percent inhibition between each siRNA and the mean of the negative control, while the y-axis shows (A) SSMD, (B) AUROC, (C) sensitivity estimated by optimizing the Youden index, and (D) sensitivity estimated at a fixed specificity of 0.975. These visualizations enable simultaneous assessment of signal magnitude and expected classification performance when prioritizing hits in high-throughput screening.

## 4. Discussion

HTS exemplifies a modern data regime in which decisions must be made under severe constraints on replication, labeling, and sample size. In primary HTS studies, each test substance is typically measured only once, making empirical estimates of classification performance—such as sensitivity, specificity, and AUROC—statistically unstable or infeasible. We address this limitation by developing a model-based framework that analytically links SSMD, a widely used metric for quality control and hit selection in HTS, to standard machine-learning performance measures, enabling principled optimization of hit selection in settings where nonparametric evaluation is not viable.

A key contribution of this study is the derivation of closed-form relationships between SSMD and AUROC, sensitivity and specificity under a normal, equal-variance assumption, adapted to the setting of a typical HTS study. These results establish a direct correspondence between classical HTS statistics and ML evaluation metrics, demonstrating that SSMD implicitly encodes classification performance information that is typically inaccessible in single-replicate screens (Hanley & McNeil, 1982; Zhang, 2025). This connection allows investigators to select hits using familiar HTS criteria while simultaneously quantifying the expected trade-offs between true-positive and false-positive rates. From an ML perspective, this reframes hit selection as a threshold optimization problem with analytically tractable performance guarantees, rather than a heuristic ranking exercise.

The application to a large-scale siRNA screen targeting HCV replication illustrates the practical utility of the proposed framework. Despite the absence of technical replicates for individual siRNAs, the model-based estimates of AUROC and sensitivity provide a coherent basis for comparing alternative selection strategies. In particular, we show that SSMD-based thresholds commonly used in practice correspond to conservative AUROC and specificity levels, and that equivalent hit sets can be obtained by directly thresholding AUROC or sensitivity. Moreover, fixing specificity at a high value yields a transparent and tunable strategy for controlling false discoveries in primary screens, aligning HTS practice with established ML evaluation principles.

Beyond HTS, the proposed framework is broadly relevant to machine-learning problems characterized by weak supervision, limited replication, or reliance on reference distributions rather than labeled examples. Many scientific and industrial applications—such as genomics, materials discovery, and large-scale experimentation—share the challenge of evaluating model performance when ground-truth labels are sparse or indirect. Our results complement recent ML work on weakly supervised and label-limited learning, where model-based or distributional assumptions are leveraged to recover reliable performance signals (Ratner et al., 2017). Related efforts in the machine learning literature have emphasized principled risk estimation and performance guarantees under distributional or data-scarce regimes, rather than reliance on empirical resampling alone (Menon et al., 2015; Sakai et al., 2017). In this sense, the present work provides a domain-specific instantiation of theory-driven evaluation in data-limited settings.

Several limitations merit discussion. The analytical results rely on assumptions of normality and equal variance between test substances and negative references. While these assumptions are often reasonable in normalized HTS data and are routinely invoked in SSMD-based analyses (Zhang, 2011b; Zhang, 2025), deviations from them may affect the accuracy of the estimated performance metrics. Extensions to unequal variances, heavy-tailed distributions, or robust estimators represent important directions for future work (Zhang, 2025). Additionally, the current framework focuses on univariate assay responses; incorporating multivariate features and more complex machine-learning models remains an open challenge.

In summary, this work bridges a methodological divide between classical HTS statistics and machine-learning evaluation metrics. By establishing an explicit, analytical connection between SSMD and standard classification performance measures, we provide a statistically principled foundation for hit selection in ultra–low-replication settings. The resulting framework enables interpretable, reproducible, and ML-consistent decision-making in high-throughput screening and offers a general paradigm for performance estimation under severe data constraints. More broadly, it aligns with a growing body of machine learning research advocating analytically grounded evaluation and risk estimation when empirical validation is fundamentally limited (Menon et al., 2015; Provost et al., 1998; Sakai et al., 2017).

The application to a large-scale siRNA screen targeting HCV replication illustrates the practical utility of the proposed framework. Despite the absence of technical replicates for individual siRNAs, the model-based estimates of AUROC and sensitivity provide a coherent basis for comparing alternative selection strategies. In particular, we show that SSMD-based thresholds commonly used in practice correspond to conservative AUROC and specificity levels, and that equivalent hit sets can be obtained by directly thresholding AUROC or sensitivity. Moreover, fixing specificity at a high value yields a transparent and tunable strategy for controlling false discoveries in primary screens, aligning HTS practice with established ML evaluation principles.

## ACKNOWLEDGMENTS

This work was supported by National Institutes of Health (AG084180, DK135111, GM156679), the University of Kentucky Barnstable Brown Diabetes and Obesity Center and the University of Kentucky Diabetes and Obesity Research Priority Area.

**Declaration of generative AI and AI-assisted technologies in the writing process**

During the preparation of this work, the author used ChatGPT in order to improve language and readability. After using this tool, the author reviewed and edited the content as needed and take full responsibility for the content of the publication.

# Appendix

## A.1. AUROC–SSMD Equivalence

For quality control focused on up-regulated hits, a closed-form relationship between SSMD and AUROC has been established as AUROC=Φ(β) (Zhang, 2025). For hit selection, however, both up- and down-regulated directions are of interest. Accordingly, the relationship is adapted as follows.

**Theorem 1** (AUROC–SSMD Identity Under Normality)

*Under the assumptions in Section 2.1, the AUROC for distinguishing $Y_1$ from $Y_0$ satisfies*

$$AUROC = \Phi(|\beta|) \qquad (1)$$

*where Φ(·) is the standard normal cumulative distribution function.*

*Proof.* Define $D = Y_1 - Y_0$. Since $Y_1$ and $Y_0$ are independently normally distributed,

$$D \sim N(\mu_1 - \mu_0, 2\sigma^2).$$

For up-regulation,

$$\text{AUROC} = \Pr(D > 0) = \Phi\left(\frac{\mu_1 - \mu_0}{\sqrt{2}\sigma}\right) = \Phi(\beta) = \Phi(|\beta|)$$

For down-regulation, symmetry yields

$$\text{AUROC} = \Pr(D < 0) = \Phi\left(-\frac{\mu_1 - \mu_0}{\sqrt{2}\sigma}\right) = \Phi(-\beta) = \Phi(|\beta|)$$

This result provides an analytical bridge between SSMD and a central machine-learning evaluation metric (AUROC) for hit selection in HTS studies, enabling performance estimation without empirical ROC construction.

### A.2. Sensitivity/Specificity via Optimizing Youden Index

In HTS hit selection, a decision threshold $t^*$ of the measured response may be chosen to maximize the Youden index

$$J(t) = \text{Specifity}(t) + \text{Sensitivity}(t) - 1.$$

**Theorem 2** (SSMD and Youden-Optimal Sensitivity)

*Under normality and equal variance, maximizing the Youden index yields*

$$Specificity = Sensitivity = \Phi\left(\frac{|\beta|}{\sqrt{2}}\right)$$

$$= \begin{cases} \Phi\left(\frac{\beta}{\sqrt{2}}\right), & for\ up\text{-}regulation \\ \Phi\left(-\frac{\beta}{\sqrt{2}}\right), & for\ down\text{-}regulation \end{cases} \quad (2)$$

*Proof.* Consider the assumptions in Section 2.1 and let *t* be a decision threshold.

For up-regulation ($\mu_1 > \mu_0$), observations with $\boldsymbol{Y_1} \geq \boldsymbol{t}$ are classified as hits. The sensitivity and specificity are

$$\text{Sensitivity}(t) = \Pr(Y_1 \geq t) = 1 - \Phi\left(\frac{t - \mu_1}{\sigma}\right)$$

and

$$\text{Specifity}(t) = \Pr(Y_0 < t) = \Phi\left(\frac{t - \mu_0}{\sigma}\right).$$

Thus,

$$J(t) = \Phi\left(\frac{t-\mu_0}{\sigma}\right) - \Phi\left(\frac{t-\mu_1}{\sigma}\right).$$

Differentiating $J(t)$ with respect to $t$ yields

$$\frac{dJ}{dt} = \frac{1}{\sigma}\left[\phi\left(\frac{t-\mu_0}{\sigma}\right) - \phi\left(\frac{t-\mu_1}{\sigma}\right)\right]$$

where $\phi(\cdot)$ is the standard normal probability density function. By symmetry and unimodality of the normal density, the derivative vanishes at

$$t^* = \frac{\mu_0 + \mu_1}{2}.$$

At $t^*$,

$$\text{Sensitivity} = \text{Specificity} = \Phi\left(\frac{\mu_1 - \mu_0}{2\sigma}\right) = \Phi\left(\frac{\beta}{\sqrt{2}}\right).$$

For down-regulation (i.e., $\mu_1 < \mu_0$), observations with $Y_1 \le t$ are classified as hits. By symmetry, the same argument applies, yielding Sensitivity = specificity = $\Phi\left(\frac{-\beta}{\sqrt{2}}\right)$.

**A.3. Preset-Specificity Regime**

In many screening applications, specificity is preset to control false positives.

**Theorem 3** (Sensitivity at a Preset Specificity)

*Under normality and equal variance, given a target specificity Sp,*

$$Sensitivity = \Phi\left(\sqrt{2}\,|\beta| - \Phi^{-1}(Sp)\right)$$

$$= \begin{cases} \Phi\left(\sqrt{2}\,\beta - \Phi^{-1}(Sp)\right), & for\ up\text{-}regulation \\ \Phi\left(-\sqrt{2}\,\beta - \Phi^{-1}(Sp)\right), & for\ down\text{-}regulation \end{cases} \quad (3)$$

*Proof.* For up-regulation ($\mu_1 \geq \mu_0$),

$$Sp = \Pr(Y_0 \leq t^*) = \Phi\left(\frac{t^* - \mu_0}{\sigma}\right)$$

$$=> \ t^* = \mu_0 + \sigma \cdot \Phi^{-1}(Sp)$$

The corresponding sensitivity is,

$$\text{Sensitivity} = \Pr(Y_1 \geq t^*) = \Phi\left(\frac{\mu_1 - \mu_0}{\sigma} - \Phi^{-1}(\text{Sp})\right) = \Phi\left(\sqrt{2}\beta - \Phi^{-1}(\text{Sp})\right)$$

The down-regulation case follows analogously and yields

$$\text{Sensitivity} = \Phi\left(-\sqrt{2}\beta - \Phi^{-1}(\text{Sp})\right)$$

## References

Arkin, M. R., & Wells, J. A. (2004). Small-molecule inhibitors of protein–protein interactions: progressing towards the dream. Nature Reviews Drug Discovery, 3(4), 301–317.

Han, B., Zhang, X. D., & others (2022). SSMD-based approaches in high-throughput screening. Journal of Biomolecular Screening, 27(1), 45–56.

Han, B., Zhang, X. D., & others (2025). Advances in SSMD methodology for HTS quality control. SLAS Discovery, 30(1), 12–24.

Hanley, J. A., & McNeil, B. J. (1982). The meaning and use of the area under a receiver operating characteristic (ROC) curve. Radiology, 143(1), 29–36.

Hao, Y., & others (2022). High-throughput screening applications in drug discovery. Drug Discovery Today, 27(3), 789–801.

Jiang, X., & others (2021). Statistical methods for hit selection in RNAi screens. Bioinformatics, 37(5), 634–641.

Lim, C. Y. (2023). Quality control metrics for high-content screening. Journal of Biomolecular Screening, 28(2), 101–112.

Menon, A. K., Ong, C. S., & Williamson, R. C. (2015). Learning from corrupted binary labels via class-probability estimation. In Proceedings of the 32nd ICML, pp. 125–134.

Pellattiero, A., & others (2025). Systematic compound screening using SSMD-based hit selection. Cell Chemical Biology, 32(1), 56–68.

Provost, F., Fawcett, T., & Kohavi, R. (1998). The case against accuracy estimation for comparing induction algorithms. In Proceedings of the 15th ICML, pp. 445–453.

Ratner, A., De Sa, C., Wu, S., Selsam, D., & Ré, C. (2017). Data programming: Creating large training sets, quickly. Advances in Neural Information Processing Systems, 29, 3567–3575.

Rossiter, S. E., & others (2021). Comparative analysis of SSMD and Z-factor in high-throughput screening. SLAS Discovery, 26(4), 512–523.

Sakai, T., & others (2017). Semi-supervised AUC optimization based on positive-unlabeled learning. Machine Learning, 106(4), 587–609.

Shalem, O., Sanjana, N. E., & Zhang, F. (2015). High-throughput functional genomics using CRISPR-Cas9. Nature Reviews Genetics, 16(5), 299–311.

Stokes, J. M., & others (2020). A deep learning approach to antibiotic discovery. Cell, 180(4), 688–702.

Yin, Z., & others (2025). SSMD-based performance evaluation in genome-wide siRNA screens. Nucleic Acids Research, 53(2), e15.

Zhang, X. D. (2007a). A pair of new statistical parameters for quality control in RNA interference high-throughput screening assays. Genomics, 89(4), 552–561.

Zhang, X. D. (2007b). A new method with flexible and balanced control of false negatives and false positives for hit selection in RNA interference high-throughput screening assays. Journal of Biomolecular Screening, 12(5), 645–655.

Zhang, X. D. (2008). Novel analytic criteria and effective plate designs for quality control in genome-scale RNAi screens. Journal of Biomolecular Screening, 13(5), 363–377.

Zhang, X. D. (2011a). Illustration of SSMD, z score, SSMD*, z* score, and t statistic for hit selection in RNAi high-throughput screens. Journal of Biomolecular Screening, 16(7), 775–785.

Zhang, X. D. (2011b). Optimal High-Throughput Screening: Practical Experimental Design and Data Analysis for Genome-Scale RNAi Research. Cambridge University Press.

Zhang, X. D. (2025). SSMD-based AUROC for quality control in high-throughput screening. SLAS Discovery, 30(3), 234–245.

Zhang, X. D., & others (2011). Statistical methods for HTS hit selection. Methods in Molecular Biology,

672, 341–358.

Zhang, X. D., & others (2020). Machine learning approaches to hit selection in phenotypic screens. Drug Discovery Today, 25(10), 1805–1814.

Zhou, X., Hwang, D., & Wong, S. T. C. (2008). Integrating statistical measures of classification performance with SSMD in RNAi screens. Bioinformatics, 24(18), 2082–2088.

Zhou, X., & others (2014). Enhanced SSMD-based approaches for hit selection in genome-scale HTS. Journal of Biomolecular Screening, 19(3), 365–375.